\documentclass[final,3p,times,twocolumn]{elsarticle}

\usepackage{amssymb}
\usepackage{amsmath}
\usepackage{graphics}
\usepackage{caption}
\usepackage{subcaption}
\numberwithin{equation}{section}

\usepackage{natbib}
\renewcommand\harvardurl[1]{\textbf{URL:} \url{#1}}
\usepackage{hyperref}

\journal{Advances in Space Research}

\begin{document}

\begin{frontmatter}



\title{A Minimal Model for Understanding Secondary Cosmic Rays}


\author[inst1]{Ramanath Cowsik}

\affiliation[inst1]{organization={Washington University in Saint Louis, Department of Physics \& McDonnell Center for the Space Sciences},
            addressline={One Brookings Drive}, 
            city={Saint Louis},
            postcode={63130}, 
            state={Missouri},
            country={United States of America}}

\author[inst1]{Dawson Huth}

\begin{abstract}
We take a phenomenological approach in a minimal model to understand the spectral intensity of secondary cosmic-ray particles like positrons, antiprotons, Lithium, Beryllium and Boron. Our analysis shows that cosmic rays at $\sim$ GeV energies pass through a significant amount of matter in regions surrounding the sources. This grammage decreases with increasing cosmic-ray energy and becomes negligible beyond $\sim 100$ GeV. During the subsequent propagation in the interstellar medium cosmic rays of all energies up to $\sim 10^5$ GeV/$n$ pass through about 1-2 g cm$^{-2}$ of matter before leaking into the intergalactic medium. It is in the interstellar medium that the bulk of the positrons and antiprotons are generated. Also cosmic-ray nuclei like C, N, and O at all energies generate additional amounts of Li, Be and B nuclei with a spectrum similar to those of C, O etc. The implications of these findings of the minimal model to the observations of gamma rays and also the importance of spatial and temporal discreteness of cosmic-ray sources for modeling cosmic-ray propagation are briefly pointed out.
\end{abstract}


\end{frontmatter}


\section{Introduction}

In recent years a large number of theoretical models have been described to understand the origins, propagation and spectral intensities of the observed cosmic rays (\cite{Strong_2007, Mertsch_2009, Mertsch_2014, Blasi_2009, Shaviv_2009, Cowsik_2009, Cowsik_2010, Cowsik_2014, Cowsik_2016, Nussinov_2016, Gaggero_2013, Evoli_2019, Evoli_2020PRL, Evoli_2020, Tjus_2020, Schroer_2021}). The main aim of the present paper is to present as briefly as possible a minimal model to understand the intensities of secondary cosmic rays. Here we focus on direct inferences that can be made from the observed intensities of various cosmic-ray components, and attempt at a parameter-free analysis as far as possible. In order to guide our phenomenological approach we begin with a broad review of the observations:

(i) Gamma-ray emission from the Galaxy: Observations in the gamma-ray band with dedicated instruments such as SAS-II, COS-B, Egret, CGRO, Fermi, Veritas, HESS, etc. have revealed the presence of a large number of high-energy gamma-ray sources in the Galaxy (\cite{SAS-II_1975, Cos-B_1977, Egret_2008, Fermi_2010, HESS_2014, Hanabata_2014, Abeysekara, Veritas_2017, Aharonian_2019, Schroer_2021, Tiblado_2021}). Besides a diffuse Galactic background in gamma-rays these discoveries have not quite fulfilled the early  expectations that this is a sure way of identifying the cosmic ray sources (\cite{Morrison_1956}). However, these observations show that cosmic rays subsequent to their acceleration interact with matter, magnetic and radiation fields in the source regions. The recent work of Blasi and his collaborators show that magnetohydrodynamic analysis indicates that such a shell or bubble formation of energetic charged particles before subsequent leakage into the interstellar medium is expected on theoretical grounds (\cite{Schroer_2021}). Eichler and Nath have pointed out that the number density of these gamma-ray sources noted above is sufficiently large so that the line of sight of the gamma-ray telescopes is likely to intersect several of these sources in the disc of the Galaxy (\cite{Nath_2020}). We will return to these theoretical ideas in the discussion section. Observations have also revealed the existence of a diffuse background of energetic gamma-rays associated with our Galaxy. The intensity of this diffuse background is such that cosmic rays at spectral intensities observed near the Earth are adequate to generate this background while interacting with the interstellar medium and radiation fields (\cite{Delahaye_2011, Ackermann_2013, Ackermann_2015, Kalashev_2016, Biswas_2020}).

The gamma-ray observations, briefly noted above, have identified two regions where the primary cosmic rays subsequent to their acceleration suffer interactions with the ambient matter and with magnetic and radiation fields: 1) in the immediate vicinity of the sources and 2) in the general interstellar medium. These observations imply that other products of the cosmic-ray interactions generated in these two regions such as antiprotons, positrons and spallogenic Li, Be, B, etc. would also be observable in cosmic rays. As the ratio of the energy of parent particles like protons to that of $\bar{p}$ and $e^+$ is very large, the contribution to the fluxes of $\bar{p}$ and $e^+$ will be suppressed if the high energy particles diffuse away from the source regions more rapidly at higher energies. Even though the Eichler-Nath study makes it difficult to assign precise ratios of the relative contributions of these sources, two distinct sources of secondary cosmic rays have been clearly identified by these observations. It appears that the contributions of these two sources to the observed intensities are of the same order of magnitude. One of the objectives of the present study is to provide an estimate of their relative contributions. 

Radiative losses modify significantly the spectral intensities of the electronic components of cosmic rays during their propagation subsequent to acceleration or production as secondaries in cosmic-ray interactions. Accordingly in the assessment of their spectra and of nuclei that suffer radioactive decay we need to keep in mind the various categories of possible sources with characteristics that are four combinations of their temporally and spatially discrete or continuous nature, each of which leave their distinctive signatures on the observed spectra.

(ii) Muons - the penetrating component of cosmic   rays: We have borrowed this phrase from Bhabha (\cite{Bhabha_1938}) a pioneer in cosmic-ray studies. The muons generated in the atmosphere through cosmic-ray interactions and arise mainly through the decay of pions. Their study dates back to the earliest days of cosmic-ray physics (\cite{Hess_1911, Hess_1912, Hess_1913, Compton_1934}). Because of their high mass and long lifetime their spectrum at high energies suffers very little modification and has provided important information on the spectral intensities of the nucleonic component of cosmic rays. Their intensity and spectrum is well-measured with magnetic spectrometers and hodoscopes located at sea level and also deep underground and under water (\cite{Allkofer_1971, Ayre_1975, Rastin_1984, Crouch_1987, Andreyev_1987, Berger_1989, DePascale_1993, Aglietta_1995, Ambrosio_1995, Bellotti_1996, Kremer_1999, Bellotti_1999, Boezio_2000, Coutu_2000, Waltham_2001, Boezio_2003, Haino_2004, Achard_2004}). Scaling of high energy interactions and straight-forward calculation of pion decay in an exponential atmosphere yields the result (for example \cite{Cowsik_1963, Cowsik_1966a, Gaisser_2016, Zyla_2020})
\begin{equation}
\begin{split}
    \frac{dN_{\mu}(E)}{dE_{mu}d\omega} = AE^{-\beta}\left\{\left(\frac{E}{\epsilon_{\pi}\eta_{\pi\mu}} + 1\right)^{-1} + \dots\right\} \\\dots \mathrm{cm}^{-2}\mathrm{s}^{-1}\mathrm{Sr}^{-1}\mathrm{GeV}^{-1}.
\end{split}
\end{equation}

Here $\epsilon_{\pi} = \frac{h_o m_{\pi}}{c \tau_{\pi}}$, $h_o$ is the scale height of the Earth's atmosphere, $\eta_{\pi \mu}$ is the weighted fraction of the energy carried away by the muons in the decay of the pion, the ellipsis inside the flower bracket represents a similar term for the contribution of kaons ($\sim 5$\%) and ellipsis outside the bracket is to account for the decay of the muons (important at $\sim$ 1 GeV here). $A$ is a constant, depending on the normalization to the cosmic-ray fluxes and various parameters like production cross-sections, interaction mean free path etc. and $\beta$ is the spectral index of the spectral intensities of nucleons in primary cosmic rays. In the 1960's $\beta$ was chosen to be 2.67 and in recent decades $\beta = 2.7$ has been the choice. The muon fluxes and the expression given in Equation 1.1 have been used widely in the study of neutrino physics (for example \cite{Cowsik_1966b, Cowsik_1999, Cowsik_2012, Dev_2016, Kajita_2010, Gaisser_2019, Rius_2017, Sui_2018}) and  for normalizing the theoretical fluxes of neutrinos. Here we use the muon spectrum as a guide in the context of analyzing the spectral intensities of secondary cosmic rays like $e^+,\ \bar{p}$, Li, Be, B etc, especially for the calculation of the spectrum of $e^+$, which is produced by the decay of the muons.

(iii) Constancy and isotropy of the cosmic-ray flux: Cosmic rays interacting with C, N and O nuclei in the Earth's atmosphere generate $^{10}$Be nuclei, which suffer $\beta$-decay with a half-life of $1.387\times10^6$ years (lifetime $\sim 2.0\times10^6$ years). These nuclei get leached by rain and get incorporated in the sea sediments. The study of the concentrations of $^{10}$Be nuclei with depth in the sediments establishes the constancy of the cosmic-ray fluxes to within $\sim10\%$ on the average over the past $\sim$10 million years. Similarly the concentrations of various cosmogenic radioactive nuclides in primitive meteorites, including $^{10}$Be and $^{40}$K, establishes the approximate constancy over much longer periods of $\sim$1 billion years (\cite{Wieler_2013}). The recent observation of $^{60}$Fe in cosmic rays (T$_{\frac{1}{2}} = 2.62\times10^{6}$ years or $\tau = 3.8\times10^{6}$ years) at the $\sim 7.5\times 10^{-5}$ level compared with $^{56}$Fe  do indicate the possibility of a small but significant contribution to the local cosmic ray flux by nearby supernovae at distances of $\sim 300 - 500$ pc from which they can propagate without significant decay. The low level of $^{60}$Fe and the other data discussed here show there has not been a major contribution from a single supernova event to the cosmic ray flux at the Earth over the past tens of millions of years (\cite{Israel_2016}).

Galactic cosmic rays diffuse into the solar system and this inward diffusion is counteracted by outward convection in the magnetized solar wind. However, diffusion becomes progressively more dominant at high energies and the cosmic-ray flux at energies greater than $\sim 300-1000$ GeV can penetrate through the solar magnetic fields and reach the Earth retaining their intrinsic anisotropy. Observations compiled by (\cite{Cowsik_2010}) indicate that the anisotropy in the fluxes is less than $\sim10^{-3}$ at energies beyond $10^3$ GeV at least up to $\sim3\times10^5$ GeV. These features show that there are no transients in the cosmic-ray flux and multiple sources contribute to the cosmic-ray flux.

(iv) Observations of spectral intensities of cosmic rays: Detailed measurements of the composition and spectra of various components of cosmic rays have been going on for a long time  with progressively better instruments (\cite{Ahn_2006, Panov_2006, Panov_2009, Chang_2008, Adriani_2014, PAMELA_2017, Abe_2016, Murphy_2016, Maestro_2019, AMS_2021}). In this brief paper we mainly focus on the data provided by the Alpha Magnetic Spectrometer and CALET because of their broad energy range and nearly complete coverage of all the charged components up to reasonably high Z values. Also, we thought it will be useful to analyze the data on various components from only the AMS and CALET instruments as the efficiency of observation will not be significantly different between neighboring charge groups and geometric factors will be similar.

We show in Figure 1 the spectra of $p,\ \bar{p},\ \mathrm{and} \  e^+$ and in Figure 2 the spectra of Li, Be, B, C, O, (Li + Be + B) and (C + O) all measured by AMS (\cite{AMS_2021}). We fit these spectra with simple power laws in the rigidity region 4 GV to 500 GV, away from rigidity domains where the spectra are not significantly affected by solar modulation or by proximity to maximum detectable momenta. We have listed in Table 1 the spectral indices of these components of cosmic rays. It is seen from Table 1 that the spectra of $p,\ \bar{p},\ \mathrm{and} \  e^+$ have very similar spectral indices (within $\pm 0.02$ of their mean value) but the spectral indices of Li, Be, and B are significantly steeper than the spectra of C and O, (steeper by $\sim0.3$ in the index). This important difference between the characteristics of the two groups with respect to the spectrum of their progenitors is addressed by the minimal model.

\begin{figure}[!t]
  \centering
  \includegraphics[width=0.45\textwidth]{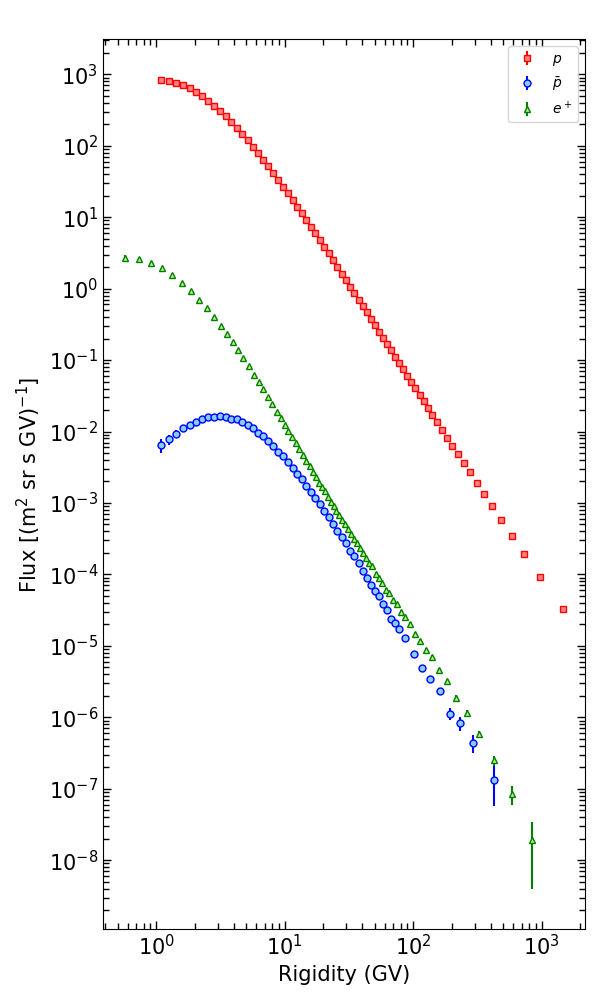}
  \captionof{figure}{AMS-02 spectra of $p,\ \bar{p},\ \mathrm{and}\ e^+$. Note that the spectral index of these three components are very close to one another, indicating that $\bar{p}$ and $e^+$ are products of high energy interactions of the abundant primary proton component of cosmic rays}
\end{figure}

\begin{figure}[!t]
  \centering
  \includegraphics[width=0.45\textwidth]{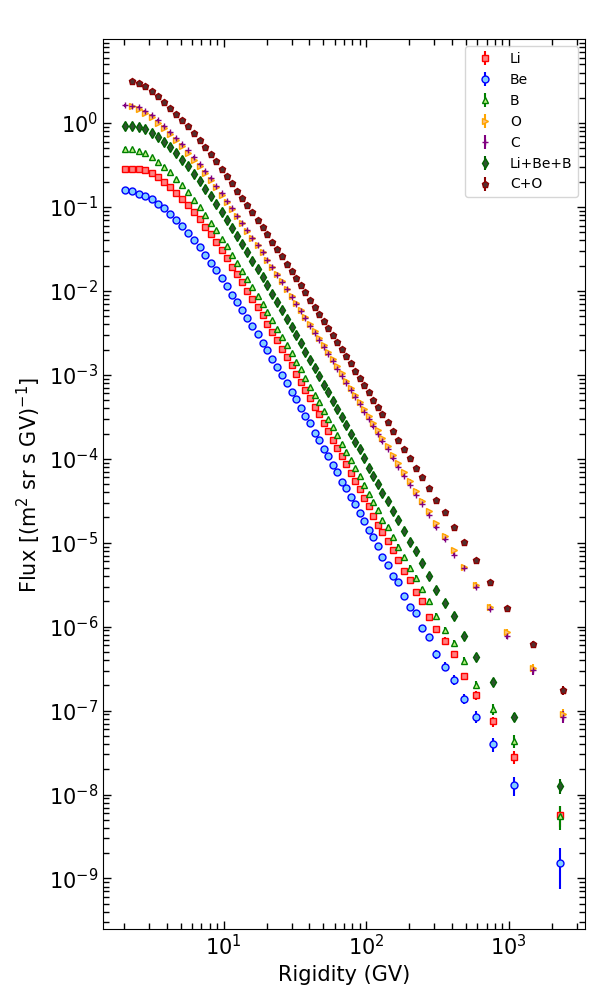}
  \captionof{figure}{The spectra of the primary C and O nuclei and those of the spallogenic Li, Be and B nuclei are shown (including the (C+O) and(Li+Be+B) spectra). Note that the spectra of Li, Be and B are significantly steeper - a feature which is addressed in the main body of this paper.}
\end{figure}

\begin{table*}[ht]
\resizebox{\textwidth}{!}{%
\def\arraystretch{1.10}
    \centering
    \begin{tabular}{c|c|c|c|c|c|c}
    \hline
    Particle & $q_o$ & $\beta$ & $b$ (GeV Myrs)$^1$ & $\tau$ (Myrs) & Fit range (GV or GeV) \\
    \hline
    \hline
    $e^+$ & $8.49 \pm 0.11$ & $2.81 \pm 0.01$ &  $1.6\times10^{-3}$ & $\sim$1 & 4 - 500 \\
    \hline
    $e^-$ & $272.65 \pm 4.24$ & $3.09 \pm 0.01$ &  $1.6\times10^{-3}$ & $\sim$1 & 8 - 500 \\
    \hline
    $p$ & $(1.62 \pm 0.01)\times10^{4}$ & $2.772 \pm 0.003$ &  - & - & 20 - 500 \\
    \hline
    $\bar{p}$ & $3.29 \pm 0.16$ & $2.77 \pm 0.02$ &  - & - & 20 - 500 \\
    \hline
    He & $(2.27 \pm 0.02)\times10^{-3}$ & $2.659 \pm 0.003$ &  - & - & 20 - 500 \\
    \hline
    Li & $29.58 \pm 0.77$ & $2.95 \pm 0.01$ &  - & - & 20 - 500 \\
    \hline
    Be & $13.21 \pm 0.22$ & $2.93 \pm 0.01$ &  - & - & 20 - 500 \\
    \hline
    B & $41.86 \pm 0.84$ & $2.96 \pm 0.01$ &  - & - & 20 - 500 \\
    \hline
    C & $63.70 \pm 0.70$ & $2.624 \pm 0.003$ &  - & - & 20 - 500 \\
    \hline
    O & $65.72 \pm 1.09$ & $2.622 \pm 0.004$ &  - & - & 30 - 500 \\
    \hline
    \end{tabular}}
    \caption{Table of best fit parameters for various cosmic-ray particles. All spectra were fit with a power law except for the $e^+$ and $e^-$ spectra which were fit using the model that includes radiative losses shown in Equation 2.}
\end{table*}

\section{The analysis of Data and the minimal model}

\subsection{Brief description of the minimal model}
Cosmic rays are generated in a large number of sources distributed in the disc of the Galaxy and subsequent to acceleration they are contained in a region surrounding the sources forming a bubble or a shell not unlike the one described by the theoretical studies of (\cite{Schroer_2021}). The escape of the cosmic rays from this bubble is dependent on energy, becoming more rapid at high energies. The subsequent propagation in the interstellar medium leading to the leakage of cosmic rays from the Galaxy is essentially independent of their energy up to $\sim 10^5$ GeV (More detailed description is to be found in (\cite{Cowsik_1973, Cowsik_1975}, \cite{Cowsik_2014} and \cite{ Cowsik_2016}). 
Beyond this energy the leakage rate increases with energy and the spectrum of cosmic rays is modeled as rigidity dependent leakage of cosmic rays injected into the Galaxy by the remnants of Type-I and Type-II supernovae by (\cite{Gaisser_2013})

\subsection{Spectral intensities of positrons}

We first address the spectral index of the positrons, as this component exemplifies nearly all the features of the minimal model. We noted that the spectral intensity of positrons in the interval $4-500$ GeV matches very well with that of the dominant proton component:
\begin{enumerate}
    \item Scaling of the pion production cross sections in high energy interactions and the subsequent $\pi^+ \rightarrow \mu^+ \rightarrow e^+$ decay ensures that the production spectrum of positrons $q_{e^+}(E)$ has the same spectral slope as that of the protons observed in primary cosmic rays.
    \item \textit{This similarity between proton and positron spectra signifies the parent-daughter relationship between the two components.}
    \item The observed positron spectrum $f_{e^+}(E)$ is just the product $q_{e^+}(E)\tau(E)$, where $\tau(E)$ is the effective residence time of the positrons in the Galaxy. Because of the scaling of the cross-section for higher energy interactions $q_{e^+}(E)$ has the same energy dependence as their parents the protons. \textit{Thus the similarity of the two spectra further implies that $\tau(E)$ is independent of energy, $E$.}
    \item The observed intensity level of $f_{e^+}(E)$ corresponds to an effective grammage, $\lambda \approx 2$ g cm$^{-2}$ for cosmic rays in the interstellar medium (\cite{Cowsik_2009, Cowsik_2010, Cowsik_2014}). \cite{Nath_2012} have suggested a similar grammage of $\sim 1-2$ g cm$^{-2}$ in their analysis of the ionization rate of the molecular clouds in the interstellar medium by cosmic rays.
    \item The radiative energy losses suffered by the positrons during their residence in the interstellar medium for a duration $\tau$, should modify their spectrum. There is indeed a noticeable and progressive steepening of the spectrum of positrons beyond $\sim$ 300 GeV corresponding to the energy losses suffered in a magnetic field of $\sim 5\ \mu$G and the cosmic microwave background at 2.7 K. At these high energies the Klein-Nishina cross sections are relevant for the scattering of positrons by starlight and such scatterings do not contribute significantly to the energy loss (\cite{Cowsik_2014}). In the analysis of the positron spectrum we may note that the source of positrons being the interactions of primary cosmic rays with the interstellar medium is both spatially and temporally continuous.
\end{enumerate}
The resulting equilibrium spectrum of positrons takes the form
\begin{alignat}{2}
    f_{e^+}(E) & = \int_{0}^{1/bE} q_oE^{-\beta}(1-bEt)^{\beta -2}e^{-t/\tau}dt\\
    & \approx q_o \tau E^{-\beta} && \mathrm{\ for\ } E\ll 1/b\tau\\
    & \approx \frac{q_o}{b(\beta - 1)}E^{-(\beta + 1)} && \mathrm{\ for\ } E\gg 1/b\tau
\end{alignat}
Here, $q_oE^{-\beta}$ is the input source spectrum of positrons per unit time into the interstellar medium, 
$b$ the radiative energy loss rate of positrons $\sim 1.6\times10^{-3}$ Myr$^{-1}$ GeV$^{-1}$ ($bE^2$ determines the actual energy loss rate) and
$\tau$ is the leakage lifetime of cosmic rays taken to be independent of energy. We display in Figure 3. $f_{e^+}(E)$ for various values of the lifetime $\tau$. From a perusal of the figure we may conclude that an energy independent lifetime $\tau \sim 1-2$ Myr fits the data, including the slight steepening seen at energies beyond $\sim$ 500 GeV. At the highest energies beyond $\sim 1000$ GeV the positron spectrum generated in the interstellar medium will reach a larger spectral index, one higher than the nucleonic component that generated it, that is $\sim \beta = -3.7$ for $E\ >$ 1000 GeV. Around $10^4$ GeV the position spectrum may steepen again reflecting the steepening of the primary spectrum beyond $\sim 10^5$ GeV/$n$.

\begin{figure}[t]
  \centering
  \includegraphics[width=0.45\textwidth]{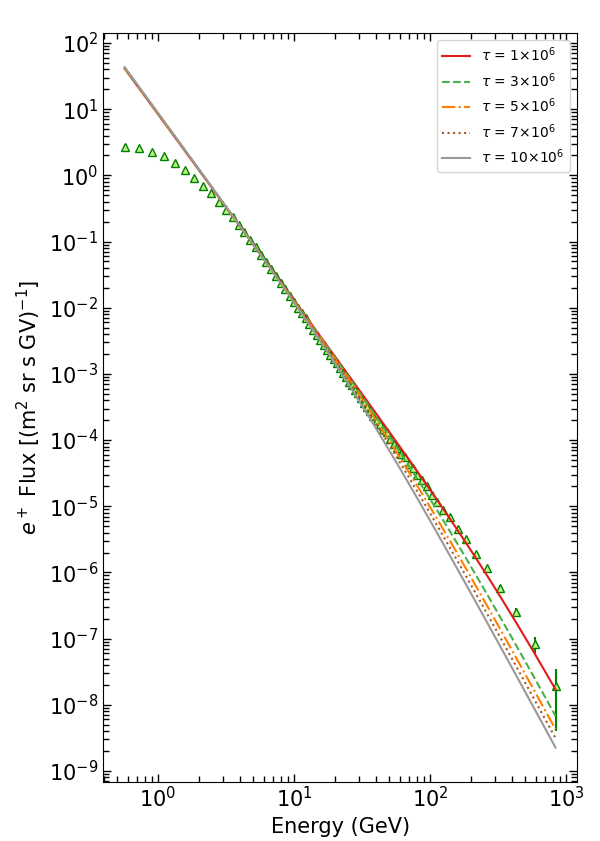}
  \captionof{figure}{AMS-02 positron spectra with various choices of $\tau$ in the model with radiative losses, all other parameters are taken from Table 1. Even though the spectral intensities at low energies will increase linearly with the assumed lifetime $\tau$, all the theoretical spectra are normalized at $\sim 10$ GeV}
\end{figure}

It may be noted that the amount of grammage needed to generate the intensities of the positrons is about 2 g cm$^{-2}$, which is equal to the grammage that will be traversed by cosmic rays moving in the interstellar medium for a period of 1-2 Myr estimated from the spectral shape. The amount of matter traversed by the cosmic rays in the interstellar medium can not exceed $\sim 2$ g/cm$^{-2}$, otherwise it will lead to spectral intensities of positions exceeding the observations.

The observation of sources of GeV gamma rays, as noted before, show that cosmic rays interact with matter in the vicinity of the sources before they diffuse into the interstellar medium. What is the level of contribution of these interactions to the positron flux?

According to the kinematics of pion production $\sim 10 \%$ of the energy of the incoming nucleon gets transferred to the leading pion, which decays transferring about $\sim 75\%$ of its energy to the muon, which in turn transfers $\sim 33\%$ of its energy to the positron. The net effect is that the weighted average energy of the proton which contributes to the production of a positron is $\sim 25$ times that of the positron, for energies beyond $\sim 5$ GeV. Thus the energy per nucleon needed even for the production of positrons of a few GeV is around $60 - 80$ GeV/$n$. In the minimal model (see for e.g. \cite{Cowsik_2014, Cowsik_2016}) at these energies the nucleonic component diffuses away from the vicinity of the sources without much interaction with the matter in the ambient medium. \textit{Thus the production kinematics ensures that positrons of $E \gtrsim$ 5 GeV get negligible contribution from the interactions of cosmic rays in matter surrounding the sources.} We will revisit some of these distinctive features induced by kinematics later in the discussion of antiprotons and spallation products.

Processes very similar to the production of positrons also generate electrons, except in this case the $\pi^-$ decay chain is involved. As the net isospin of the nuclear active component is positive more positrons are produced than electrons. Thus the positron fraction, $e^+/(e^+ + e^-)$ at production is $\sim 0.6$ (\cite{Cowsik_2009, Cowsik_2010}). This has interesting consequences for the expectation of the positron fraction at high energies. \textit{We note here that this source of positrons and electrons by cosmic-ray interactions in the interstellar medium is both spatially and temporally continuous.}

\subsection{The electronic component of cosmic rays}

Observations with instruments such as CALET do not distinguish $e^+$ from $e^-$ but yield the sum of the intensities of these two particles, which is called the electronic component. To understand this component it is useful to subtract from it the secondary electronic component generated in the interstellar medium by the nuclear component of cosmic rays - the spatially and temporally continuous source (\cite{Cowsik_2009}). The CALET data extends the earlier data beyond 1500 GeV up to $\sim 4000$ GeV and confirms with better statistics the steepening of the electronic component beyond $\sim 800$ GeV. The subtraction of the secondary electronic component generated in the interstellar medium will tend to sharpen the steepening leading to an essential cut-off in the primary electronic component beyond a few TeV. (As an aside, we note that a very early analysis with a doubly continuous source model had indicated a lifetime $\sim 1 - 2$ Myr for the electronic component (\cite{CowsikPRL_1966})) Two important points need to be made here:
\begin{enumerate}
    \item The secondary electronic component generated in the interstellar medium is spatially continuous (to the extent that the ISM is smooth) and temporally continuous (to the extent that cosmic ray intensity in the ISM is constant). Accordingly, its spectrum will steepen to a form $\sim E^{-3.7}$ and will extend up to $\sim 10^4$ GeV. Beyond this it would steepen reflecting the steepening of the cosmic ray spectrum beyond $\sim 10^6$ GeV.

    A direct consequence of this persistence of the secondary component is \textit{that at very high energies the positron fraction will tend to $\sim 0.6$ reflecting its production by the nuclear component in collisions with the interstellar matter.}
    
    \item The primary electronic component which is produced in the sources will take a finite time to diffuse from the sources to the Earth where it is detected. Earlier analysis that assumes that even though the sources are spatially discrete, they are temporally continuous (\cite{Cowsik_1979, Cowsik_2014, Nishimura_1997}) indicate a cut-off in the primary electronic component of the form 
    \begin{equation}
        f_n(E) \sim \mathrm{exp}\left[\frac{-b r_n^2 E E_x}{4 \kappa (E_x - E)}\right]
    \end{equation}
    Here $r_n$ is the distance to the $n$th nearest source, $E_x$ the maximum energy up to which the primary electronic component is accelerated, $b\sim1.6\times10^{-3}$ GeV$^{-1}$ Myr$^{-1}$ is a parameter which characterises the radiation energy losses in the form $\dot{E} = -b E^2$, and $\kappa$ is the value of the diffusion constant. For large values of $E_x$ the cut-off in the spectrum occurs at 
    \begin{equation}
    \begin{split}
        E_c \approx \frac{4\kappa}{b r_n^2} \sim 800 \mathrm{\ GeV} & \mathrm{\ for\ } \kappa \sim 10^{28} \mathrm{cm}^2 \mathrm{\ s}^{-1} \\
        & \mathrm{\ and\ } r_n \sim 300 \mathrm{pc}
    \end{split}
    \end{equation}
    
    If the source is transient and generated a burst  $\tau$ years ago at a distance $r_n$ then it would take a finite amount of time for the particles to diffuse the intervening distance in significant numbers.
    \begin{equation}
        \tau_n \sim \frac{r_n^2}{6\kappa}
    \end{equation}
    and the spectrum would be cut off at $E_c \approx 1/b\tau$ for large $E_x$. Furthermore, there will be considerable dilution of the flux during the further time evolution subsequent to the arrival of the diffusion front (\cite{Cowsik_2009}).
\end{enumerate}

\subsection{The antiproton component}

\textit{It is remarkable that the antiproton component, at energies $E \gtrsim 10$ GeV has the same spectral slope as the positrons and also the protons (that are the dominant particles in the primary cosmic-ray spectrum). This strongly endorses a joint origin for $e^+$ and $\bar{p}$, namely, as the secondary particles generated in the collisions of primary cosmic rays in the interstellar medium.} The kinematics of the production of antiprotons are different from those of the positrons. The threshold for $\bar{p}$ production in $p-p$ collisions is $\sim 7 m_p \sim 6.6$ GeV and the antiproton emerges even at threshold with a total energy of $\sim 1.7$ GeV. Accordingly cosmic-ray interactions in matter surrounding the sources contribute a small but significant flux at energies below $\sim 10$ GeV.

Antiprotons in cosmic rays suffer negligible amounts of annihilation in the interstellar medium and no losses of energy through radiation. Accordingly their spectrum is not modified during their lifetime in the Galaxy. Beyond $\sim$ 5-10 GeV their intensities are consistent with their residence in the Galaxy for $\sim$ 1-2 Myr.

\subsection{Spectral intensities of L and M nuclei}
The grammage of $\sim 2$ g cm$^{-2}$ in the interstellar medium estimated to generate the observed positron and antiproton intensities will concomitantly lead to the production of Li, Be and B nuclei through the spallation of C, O and other nuclei. Marrocchesi et al. of the CALET collaboration (\cite{Marrocchesi_2021}) compare the flux of B measured by them over a 5 year period with those by AMS and others. They confirm the progressive flattening of the Boron spectrum beyond $\sim 200$ GeV/$n$ and extend it to $\sim 1500$ GeV/$n$; in this energy domain we note the spectrum fits $\sim E^{-2.7}$ dependence, while at energies below $\sim 100$ GeV the spectrum is steeper, $\sim E^{-3.1}$. 

The flat portion of the Boron spectrum could be attributed to the production of these nuclei in the interstellar medium and storage for a time $\sim 2$ Myr. Extending the $E^{-2.7}$ spectrum to lower energies, we may note that it contributes to $\sim 25\%$ of the observed Boron spectrum at $\sim 10$ GeV/$n$. The difference of $\sim 75\%$ is an estimate of the contribution of the spallation in the regions surrounding the sources. The figure 9. of (\cite{Marrocchesi_2021}) displays the observed B/C ratio up to 1500 GeV/nucleon, and illustrates this point very well. If we draw a horizontal line through the point at $\sim 200$ GeV/nucleon, it reproduces the flux ratio at high energies and contributes about $30\%$ of the ratio at 10 GeV/nucleon. The difference drops steeply with increasing energy and becomes negligible beyond $\sim 100$ GeV/nucleon. In the minimal model this difference represents the energy-dependent grammage traversed by the cosmic rays in the regions surrounding the sources. The evidence for such a clear separation of the two components in the spectra of secondary nuclei is not as clear in the spectra of Lithium and Beryllium, but the AMS and CALET data do not preclude such a separation.

\subsection{The injection spectrum}

In the minimal model cosmic rays are injected with the same spectral characteristics as the observed spectra of protons and other nuclei $\sim$ E$^{-2.7}$. The spectra of gamma rays and non-thermal radio waves from young supernova remnants indicate the presence of energetic particles with flatter spectra $\sim$ E$^{-2}$ – E$^{-2.5}$ in apparent contradiction with the requirements. This issue has not been resolved definitively; however there are many possibilities: V\"{o}lk and collaborators have argued that the spectra of young supernova remnants need not necessarily represent the spectrum of particles that are finally injected into the interstellar medium (\cite{Berezhko_2006}). The analysis of the evolution and structure of supernova remnants (\cite{Brose_2021}) clearly indicates the softening of the spectrum with age with synchrotron spectrum reaching an index of $\sim - 0.75$ in $\sim 10^4$ years and suggests the possibility that by about $10^5$ years when the energetic particles completely diffuse and merge into the interstellar medium it is possible that the spectral softening will reach the required value of $– 0.85$. These particles responsible for the synchrotron emission are generally attributed to those accelerated in the reverse shock of the remnant. The acceleration in the forward shock could have different characteristics; softening of the shock profile due to the presence of neutrals, back reaction of the accelerated particles or their loss from the system due to leakage will lead to the production of a steeper spectrum. There may also be other possibilities for generating the requisite injection spectrum. At present we need to treat it as a requirement of the model – the injected spectrum of the cosmic rays ought to have the same spectral index as the observed spectrum of primary nucleonic components up to $\sim 10^5$ GeV.

\subsection{The residence time of cosmic rays in the Galaxy}
In principle there are three possibilities for achieving energy-independent leakage of cosmic rays from the Galaxy: (1) The loss of cosmic rays from the galaxy is dominated by convection or advection; (\cite{Reichherzer_2022}) invoke this mechanism to fit the observations of gamma-rays in the central regions of the Galaxy up to a radial distance of $\sim 5$ kpc. (2) Parker instability occurs in the galactic plane and causes magnetic flux tubes to emerge out of the Galactic plane inflated by cosmic-ray pressure. As this happens the interstellar gas slides back to the disk forming clouds and the cosmic rays flow out of the galaxy (\cite{Parker_1966, Parker_1967, Parker_1969, Parker_2019}). (3) The turbulence spectrum has a slope of $-2$, so that the quasilinear theory predicts a diffusion constant independent of energy; keeping in mind the findings of Reichherzer et al. this requires that the fluctuation fields have a very small amplitude.

\subsection{The anisotropy of cosmic rays}

The diffusive current of cosmic rays generates anisotropy in the local interstellar medium of magnitude $\delta \sim 3 \kappa \Delta n/cn$, where $n$ is the density of cosmic rays. The lifetime of cosmic rays gives us an estimate for the value of the diffusion coefficient $\tau \sim s^2/4 \kappa$, $s$ being the scale length of the containment volume of cosmic rays. Combining these two the anisotropy is seen to be $\sim 3 \kappa \Delta n/cn \sim 3 s/4 c \tau$, independent of energy in the minimal model, small and consistent with the observational upper bounds. It should be noted that at the Earth the anisotropies are noticeable only at energies beyond about $300\ –\ 1000$ GeV, because at lower energies the diffusive current of cosmic rays into the solar system is counteracted by the outward directed convection current in the solar wind. At high energies diffusion dominates and the interstellar anisotropies become measurable at the Earth.

\section{Discussion}

It is perhaps useful to begin the discussion as suggested by the referees with a brief review of alternate views regarding the same aspects of Galactic cosmic rays as those discussed in this paper. Before discussing the models, we note that the AMS - Collaboration favor the view that their measurements of the positron fluxes and the positron fraction imply that the positrons arise from the annihilation of Galactic dark matter particles or from pulsars (\cite{Accardo_2014, Bergstrom_2013, Linden_2013, Feng_2014, Cholis_2018}). The alternate view generally assumes diffusive transport of cosmic rays leading to their loss from the Galaxy into the intergalactic medium with a rate that increases with the rigidity or energy of the particles, thereby reducing their lifetime with increasing energy (\textit{hereafter referred to as the 'energy-dependent leakage model'}). By an appropriate choice of this energy dependence the declining ratio of the intensities of secondary cosmic-ray nuclei like Boron to the intensities of their parents like Carbon can be reproduced. In order that the spectra of primary cosmic rays like protons, helium and carbon match the observations, the sources of cosmic rays are assumed to inject a relatively flat spectrum into the interstellar medium to compensate for the enhanced leakage from the Galaxy at high energies. This alternate view was advanced nearly 50 years ago soon after the “nested leaky-box model” was proposed by (\cite{Cowsik_1973, Cowsik_1975}), which is the minimal model presented in this paper. In an early review of the status of the field (\cite{Cesarsky_1980}) remarked in conclusion that the measurement of the positron spectrum will dictate the definite choice between the two models. Today we have good measurements of the spectrum of positrons, as well as those of antiprotons, which are both similar to that of the protons and as we have shown are consistent with the predictions of the nested leaky box model.

The energy-dependent lifetime model evolved with the inclusion of many details including the possible distribution of sources and parameters pertaining to the Galaxy, and also convection and other transport mechanisms, all coded into the software called “GALPROP”. This provided considerable flexibility to vary the large number of parameters easily and explore the consequences. These rapid developments in the field were reviewed by (\cite{Strong_2007}).

Subsequently new sets of extensive observations of cosmic rays, especially space-borne detectors like PAMELA, AMS and CALET stimulated a large body of elaborate work with the participation of a great many scientists investigating even more complex sets of scenarios with computer simulations and new software packages.

The motivation for the energy-dependent lifetime models were in part provided by a wide variety of observations of the interstellar magnetic fields: These observations, mainly using radio-astronomical techniques indicate that the smooth magnetic field of the Galaxy is superimposed by random magnetic fields of about the same or even greater strength (\cite{Armstrong_1995, Iacobelli_2013}) and in the interplanetary space (\cite{Goldstein_1995, Luca_2016}).The power-law index of the wavenumber spectrum, $\alpha$ varies over a wide range from about $-2$ to $-5/3$ (Kolmogorov) or even down $-3/2$ (Kraichnan). The quasi-linear theory of particle transport predicts a diffusion constant $\kappa$ that increases with rigidity as R$^{2-\alpha}$ and the assumption of the Kolmogorov value $\alpha$ = 5/3 which immediately helps to  understand the falling (B/C) ratio in cosmic rays. The relatively flat spectra predicted by the highly efficient acceleration of cosmic rays in high Mach number shocks, and the observations of the spectra of gamma rays from several supernova remnants. If the injection spectra of cosmic rays correspond to these theoretical expectations and observations, then the energy-dependent lifetime would generate a steeper spectrum in the steady state in conformity with the observations. This is just a very brief summary of a field which incorporates many subtle aspects of physics and astrophysics reviewed by (\cite{Cesarsky_1980, Strong_2007}) and by (\cite{Tjus_2020}) and also referred to in context in this paper (\cite{Chang_2008, Cowsik_2010, Mertsch_2009, Mertsch_2014, Evoli_2019, Evoli_2020, Blasi_2021}). In such a scenario the spectra of positrons and antiprotons tend to be noticeably steeper than that of their parent particles, the protons and the other nuclei in cosmic rays. Accordingly, several papers have discussed alternate sources for these particles, sources such as annihilating Galactic dark matter and positrons diffusing out of pulsar magnetospheres (\cite{Accardo_2014, Bergstrom_2013, Linden_2013, Feng_2014}). The review by JB Tjus and L Merten provides a good overview with a more extensive reference list. However, these models do not provide a natural explanation for the observation that the spectra of both positrons and antiprotons have the same spectral index and their intensities are both compatible with a traversal of $1 - 2$ g cm$^{-2}$ of interstellar matter by the primary cosmic rays.

Two papers that have appeared recently prompt us to reconsider this picture: In the first paper by (\cite{Schroer_2021} shows that cosmic rays streaming out of the sources generate various modes and that these modes scatter cosmic rays very efficiently reducing their diffusion. Accordingly, cosmic rays are confined to a bubble or a shell surrounding the sources for sufficient time to significantly affect the ratios of secondary to primary cosmic rays like B/C. In summary they state “Our results … provide a physical explanation of the numerous claims of suppressed cosmic-ray diffusion around Galactic sources…”.  Thus the analysis of cosmic-ray phenomena within the framework of the energy-dependent leaky-box model will require reconsideration. The second paper by (\cite{Reichherzer_2022}) note that even though Kolmogorov spectrum is an idealization and the interstellar turbulence is driven on many scales from scales of $\sim$100 pc of super bubbles down to the kinetic scales of cosmic rays themselves. However, they adopt a Kolomogorov spectrum in their explicit simulations of the transport based on Taylor-Green-Kubo formalism. One of the results of their analysis is that the diffusion constant and its dependence on particle rigidity depends not only on the exponent $\alpha$ of the power spectrum, but also on the ratio of the strength of the random magnetic fields $b$ to the smooth fields $B$ in the interstellar medium. They note that the index $\delta$ depicting the increase of the diffusion constant with rigidity does not quite reach the value $\delta = 2 - \alpha$, ($\delta$ = 1/3 for Kolmogorov), predicted by the quasilinear theory until the ratio $b/B \leq 0.05$, and increases with increasing value of the ratio. On the other hand, the radio astronomical observations yield a ratio $b/B \sim 1$. This again points to the need to reexamine the energy-dependent leakage models.

We are not aware of any recent paper that has critically reviewed these recent developments, nor has a consensus view emerged regarding the parameters of the energy-dependent leakage models. These energy-dependent lifetime models are able to fit the observed positron spectrum assuming that that this arises from leakage of positrons from the magnetospheres of pulsars. However, they have not paid equal attention to explain the spectrum of antiprotons.

The extensive data collected with the AMS and CALET instruments provide good coverage of the spectra with good statistics up to $\sim 5\times10^4$ GeV for protons and up to $\sim 1.5\times10^3$ GeV/$n$ for the C and O nuclei. These primary cosmic-ray nuclei are most relevant for the generation of the secondary cosmic rays discussed in this paper. The data indicate breaks in the spectra with the spectral indices changing by $\sim 0.1$ up to $\sim 0.4$ in a narrow interval $\Delta E$ at various energies $E$. A key feature of these breaks is that they occur over a very narrow range of energies $\sim 0.5E$ to $2E$

There have been suggestions that because of the multiplicity and discrete nature of the sources such breaks are to be expected when spectra of slightly different spectral indices from a few sources overlap on top of one another. However, such a superposition, say of two power laws, would only result in a slow flattening over an energy band $\sim 0.2 E$ to $5E$

With regard to the alternate possibility that the changes could be associated with the acceleration process, it may be noted that for even in the first order diffusive Fermi acceleration the energy gain is by a factor $(1 + \frac{\nu_s}{c}) \approx 1 + 2\times 10^{-2}$ at each crossing of the shocks, and changes as sharp as indicated by the data are not expected.

Protons dominate the nuclear component of cosmic rays, which contain about 99\% of the energy density of cosmic rays, $\sim 10^{-12}$ erg cm$^{-3}$. The only Galactic source that has been identified that contains adequate power and can in principle generate this component is the supernova. The debris of the explosion carries typically $10^{50} - 10^{51}$ erg and higher values in the explosion of very massive stars, the rate of such events being about one in 50-100 years. The efficiency of transfer of kinetic energy of the debris to relativistic particles, through acceleration in shocks for example, has to be very high $\gtrsim 10\%$ in order to replenish the cosmic rays which steadily leak out of the Galaxy by diffusion.

The presence of various isotopes of Fe, Co, Ni and r-process nuclei in cosmic rays clearly point to a supernova connection. For example $^{59}$Co is produced by the electron capture of $^{59}$Ni, which in turn is the decay product of $^{59}$Cu. These nuclei are generated in the very final stages of evolution in the cores of massive stars, lasting less than a few days just before the explosion. At the same time, the observation that $^{59}$Ni has decayed by electron capture to generate the flux of $^{59}$Co in cosmic rays indicates that supernovae should accelerate the products of nucleosynthesis after a period of about 10$^{5}$ years, the half-life of $^{59}$Ni. On the other hand the observation $^{60}$Fe nuclei in cosmic rays indicate that the cosmic rays that are observed at the Earth went through the acceleration process and then diffused to the Earth within one or two lifetimes for the decay of $^{60}$Fe, about 3.7 Myr. The observations of these isotopes therefore confines the acceleration process in the interval between $10^5 - 4 \times 10^6$ ergs after the supernova explosion (\cite{Casse_1983, Wiedenbeck_1999, Binns_2014, Israel_2016}). 

In our present discussion we have not included these and other constraints posed by the observations of composition, anisotropy, etc., and focused on the implications of the observations of high energy cosmic rays by CALET and AMS. In particular we have studied the implication of the positron and antiprotons being the secondary particles generated by the primary nuclear component of cosmic rays.

(i) In the minimal model we have described here the focus has been on the ratios of the spectral intensities of secondary cosmic rays to that of their progenitors - the primary cosmic rays. The reason for this is that absolute intensities especially over a wide range of energies are very difficult to measure. This is particularly true for measurements carried out with balloon borne and satellite borne instruments. Cosmic ray studies have given birth to the whole field of high-energy astrophysics and although major effort has been dedicated to their study over the past century, the challenges posed by steep spectra and declining abundances with increasing atomic number and energy have prevented precise measurements of the spectral intensities of various components of cosmic rays. This is true even after the major progress achieved recently with instruments like AMS, ATIC and PAMELA (\cite{AMS_2021, Ahn_2006, Panov_2006, Panov_2009, Chang_2008, Adriani_2014, PAMELA_2017, Maestro_2019, Marrocchesi_2021}). In this context the spectral intensities of charge ratios of muons measured at sea level provide an insight into the spectrum and composition of the nucleonic component at high energies, especially for the analysis of $e^+$ and $\bar{p}$ in cosmic rays.

(ii) The close similarity of the spectra of $p,\ \bar{p}\ \mathrm{and}\ e^+$ also indicates that their lifetimes are independent of energy and are nearly the same in the Galaxy, as expected for scattering by fluctuations and irregularities in the magnetic fields being the controlling factor in their diffusion. Continued effort towards measurement of their spectra should be a priority for the future.

(iii) The question proposed by $^{10}$Be/$^9$Be is interesting, as the bulk of Be production in the minimal model takes place in and around the source regions. As the particles diffuse in the interstellar medium it takes a finite amount of time for them to reach us even from the nearest of the sources. Radioactive decay will significantly depopulate the $^{10}$Be nuclei with respect to $^9$Be as discussed by (\cite{ Cowsik_2014, Cowsik_2016}). It will be useful to confirm their suggestion with more recent data.

(iv) Returning to the confinement and diffusion of cosmic rays close to their sources, we may look at the recent theoretical study of (\cite{Blasi_2021}) and (\cite{Schroer_2021}) who show that the dynamical effects of cosmic rays in and around the sources confine them in these regions for sufficient times to interact with ambient matter and radiation fields. These analyses have confirmed earlier kinematic analyses and provide theoretical support for understanding gamma-ray sources and of course cosmic ray propagation.

\section{Conclusion}
The critical review of cosmic ray data implies that cosmic rays spend significant time in and around the sources generating gamma-rays, and secondary nuclei like Li, Be and B up to $\sim 300$ GeV/$n$. Some production of these secondary nuclei, and the bulk of the production of $e^+$ and $\bar{p}$, especially at high energies, occur in the general interstellar medium. The lifetime of cosmic rays in the Galaxy is nearly constant or weakly dependent on energy up to $\sim$ PeV energies, and helps us to understand the constancy and isotropy of the observed cosmic rays.

Thus the preponderance of evidence from observations of $\gamma$-ray astronomy and the spectra of positrons, antiprotons and secondary nuclei indicates that there are two regions that contribute to the secondary component in cosmic rays: (i) The region surrounding the sources dominate at lower energies and tapers off with increasing energy. (ii) The second region is the general interstellar medium contributing nearly a constant amount of grammage at about 20\% of the maximum at all energies.

\section{Acknowledgements}
We would like to thank our past collaborators B. Burch, M.A. Lee, and T. Madziwa-Nussinov for their efforts in progressing our understanding of the material in this paper. We also thank the reviewers for their carfeul reading of the paper and productive suggestions.
\begingroup
\raggedright
\bibliographystyle{elsarticle-num} 
\bibliography{refs}
\endgroup





\end{document}